# Signal to Noise Ratio optimization for extended sources with a new kind of MURA masks

I. Kaissas[a,][*], C. Papadimitropoulos[a], A. Clouvas[b], C. Potiriadis[a], and C.P. Lambropoulos[c]

[a] *Greek Atomic Energy Commission,*
*Patriarchou Grigoriou & Neapoleos, 15341 Agia Paraskevi, Attiki, Greece*

[b] *School of Electrical & Computer Engineering, Aristotle University of Thessaloniki,*
*University Campus, 54124 Thessaloniki, Greece*

[c] *National and Kapodistrian University of Athens,*
*Athens, Greece*

*E-mail*: ikaisas@gmail.com

ABSTRACT: Using coded aperture, for localization of radioactive hot-spots, results in enhanced efficiency and under certain configurations wide Field of View (FOV). We present a coded aperture assembly technique which can be restructured easily, as well as the reduction of the intrinsic noise of coded apertures constructed with this technique, when they localize spatially extended γ-emitters. Specifically, Modified-Uniformly-Redundant-Array (MURA) coded apertures are structured by embedding lead spheres in a matrix of positions machined inside a transparent medium such as acrylic glass, resulting in an advantageous transparent to opaque area ratio and thus an improved detection efficiency. This configuration also induces a systematic, element-wise, noise on the Point-Spread-Function (PSF) of the correlation matrix. When imaging with these apertures extended hot-spots, a penumbra phenomenon occurrs and reduces this intrinsic noise, in the way a kernel filter does. Fast-Fourier-Transform (FFT) is used to analyze the effect of this phenomenon on the correlation matrix and to explain the maximization of its Signal-to-Noise Ratio (SNR) for certain extent of the hot-spots. Simulations have been used for the detailed study of the SNR dependence on the dimensions of the hot-spot, while experiments with two $^{99m}$Tc cylindrical sources with 11mm and 24mm diameter, respectively and 1.5 MBq activity each, confirm the reduction of the intrinsic noise. The results define the way of optimization of the imaging setup for the detection of extended hot-spots. Such an optimization could be useful for example in the case of lymph nodes or thyroid remnant imaging in nuclear medicine. Finally, we propose a kernel filter, derived by the Auto-Correlation-Function (ACF), to be applied on PSFs with high intrinsic noise, in order to eliminate it.

---

[*] Corresponding author.

# Contents

## 1. Introduction

Coded aperture imaging could be considered as an evolution of pin-hole imaging. Instead of a single pin-hole on an opaque plate, several holes contribute to the brightness and the contrast of the γ-emitters imaging. The Non-Two-Holes-Touching (NTHT) mask [1] is a special case of the Modified-Uniformly-Redundant-Array (MURA) [2] type of apertures. If the opaque plate with holes is replaced by a transparent plate with embedded opaque elements, the detection efficiency is increased. This interchange between the opaque plate and the holes results in a mask which can be seen in Figure 1. We call this kind of mask a Non-Two Obscurations –Touching (NTOT) one. The shadowgram of the γ-emitters placed within the Field-Of-View (FOV) (either point sources or extended hot-spots), formed via the mask on the pixel detector, can be seen in Figure 2. In case of localization of one point source the resulting correlation matrix contains a systematic, element-wise noise that lays on the surrounding background of the Point Source Spread Function (PSF). Figure 2 shows the details of its shape. Its small peaks and valleys have diameter proportional to the size of the mask-element.

    Many researchers contributed to the evolution of coded apertures in near field applications. Among them Cannon and Fenimore [3], Accorsi and Lanza [4] and Mu and Liu [5] studied the artifacts that the near-field detection geometry induces. In previous works [6] [7], we developed a system of two γ-cameras (Figure 1) that exploits the parallax phenomenon and combines the two derived directions with a triangulation method, in order to find the 3D coordinates of the radioactive sources placed in the Fully Coded Field of View (FCFOV) of the system. We studied the figures of merit for near-field detection via simulations and experiments. In the current work we focus on the systematic, element-wise noise appearing on the correlation matrix, that we call intrinsic noise of the mask pattern. The rest of the background noise is defined as random noise.

    We provide the explanation of the origin of the intrinsic noise. We find the optimum combination of mask-element size and source-detector distance, which minimize the intrinsic noise for extended hot-spots with certain dimensions and we apply an image filtering method that reduces the intrinsic noise, when a point source is localized.

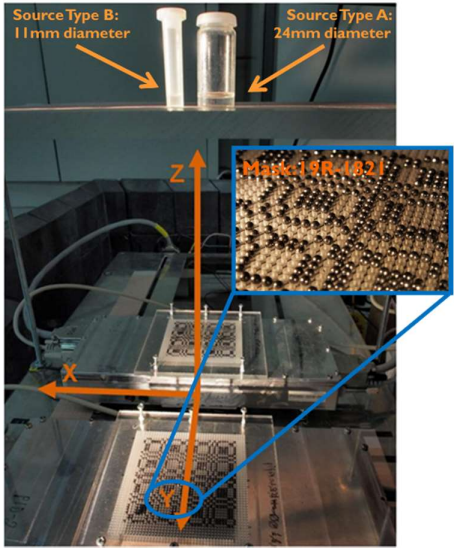
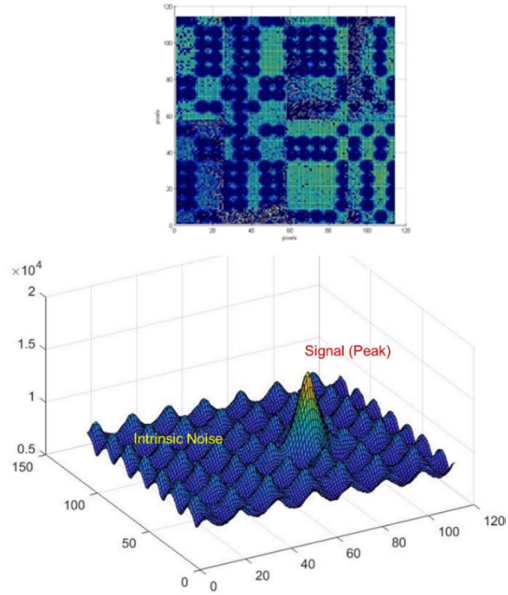

**Figure 1.** The two CdTe γ-cameras and the two types of 99mTc hot spots with different diameter. In blue frame: A close-up detail of NTOT MURA mask. Mask elements are Pb spheres arranged on a transparent acrylic glass plate.

**Figure 2.** Top: A shadowgram formed on the γ-camera by a point source. Bottom: A correlation matrix with the characteristic peaks and lows of the intrinsic, element-wise noise. The tall peak is the PSF and its diameter along with the diameter of the peaks and valleys of the intrinsic noise are proportional to the diameter of the mask elements.

Thus, a study of the Signal-to-Noise-Ratio (SNR) of the correlation matrix is mainly presented. Furthermore the standard deviation and the shape of the background of the correlation matrix are considered in the discussion of the experimental results.

## 2. Materials and Methods

The γ-camera has an active area of 44 x 44 mm$^2$ consisting of 8 hybrids. The hybrid pixel detectors are CMOS ASICs with 1mm thick CdTe crystals bump bonded. The pixel pitch is 350μm and 27 frames/sec are recorded. The energy resolution is 3-4 keV FWHM for the $^{99m}$Tc photo-peak. On top of the γ-camera a NTOT MURA mask is placed parallel to the CdTe surface at 2cm distance from it. Two different types of MURA masks were constructed with purpose to localize hot-spots at short and medium distances from the γ-camera. Table 1 presents their specifications. Their basic pattern consists of 19x19 elements and they differ basically in the element pitch (i.e. the interval between two neighbour elements), which determines the ideal source distance. This is the source-detector distance for which a point source produces the PSF on the correlation matrix without side-lobs. For each of the two masks, the corresponding ideal source distance was used.

| Mask type | Number of elements | Surface (mm$^2$) | Element pitch (mm) | Ideal source distance (mm) |
|---|---|---|---|---|
| 19R-1821 | 37×37 | 67.4×67.4 | 1.821 | 160 |
| 19R-1958 | 37×37 | 72.5×72.5 | 1.958 | 308 |

**Table 1.** Specifications of the MURA masks used in this study.

A code for fast simulation of the coded aperture imaging system has been developed and its results have been verified experimentally [7]. It is used in order to produce shadowgrams of simulated hot-spots with various spatial distributions as presented in Figure 3. Specifically, the first part of the code generates a hot-spot of uniform or normal spatial distribution with dimensions from 0 mm to 50 mm or 85 mm respectively. The photons emitted by the hot-spots which pass through the mask are counted in order to form the shadowgram. The second part of the code implements the correlation of the shadowgram with a digital form of the mask, the G matrix [2], and produces the correlation matrix. The surrounding background of the peak includes the intrinsic noise. One of the derived parameters of the correlation matrix is the SNR that is defined as the peak height minus the mean value of the background divided by the standard deviation of the background:

$$SNR = \frac{peak - mean\ value\ of\ background}{standard\ deviation\ of\ the\ background} \quad (1)$$

The simulation experiment was repeated five times, in order to obtain enough statistics, for every incremental step of the hot-spot dimension.

The second part of the code can be fed also with shadowgrams captured by the experimental setup. Two cylindrical hot-spots of $^{99m}$Tc, with activity almost 1.5MBq, were placed in the FCFOV for 184 sec each. They differed basically in their diameter, that was 24mm for source Type A and 11mm for source Type B. Both cylinders had a height 9mm. The derived correlation matrices are presented in Figure 4. The deviations in the activity and the positioning of the hot-spots slip into the SNR, preventing the verification of the simulation results in the case of small changes of the SNR. For these cases an alternative and more simplified parameter is used, the standard deviation of the background of the correlation matrix.

## 3. Results and Discussion

### 3.1 The intrinsic noise

The diameter of the intrinsic noise structure (peaks and valleys in Figure 2) is proportional to the diameter of the projection of the mask elements on the pixel CdTe detector and consequently it is also proportional to the FWHM of the PSF of the correlation matrix. As the element pitch is increased, the wavelength of the periodic structures of the shadowgram also increases. The SNR of the correlation matrix becomes maximum (Figure 3, SNR Plot) for a certain size of extended hot-spot, for which the penumbra blurs the shade of each single mask element. As far as the extension of the hot-spot eliminates, via the penumbra, the separate projection of each mask element on the shadowgram (It can be seen in the comparison between the left and the middle shadowgram of Figure 3), but it does not deteriorate the basic pattern of the mask (this deterioration can be seen in the comparison between the middle and the right shadowgram of Figure 3), the SNR of the correlation matrix is improved. A further extension of the hot-spot dimension deteriorates the basic pattern on the shadowgram, by hiding the details in the middle range of the spatial frequencies, which are essential for a good correlation with the G matrix.

The 2D FFT of the shadowgram shows clearly the elimination of the high spatial frequencies of each mask element projection. Further extension of the hot-spot eliminates also the medium spatial frequencies of the basic pattern of the shadowgram and consequently deteriorates the SNR.

A detailed comparison of experimental data with simulation is difficult, due to the deviations in the activity and the positioning of the hot-spots. However, the quite large difference in the SNR of the shadowgrams recorded from two different types of sources with the mask 19R-1821 is in agreement at least qualitatively with the conclusions drawn from the analysis of the simulation results. The type B source has a radius of 5.5 mm and consequently lays on the region of the peak SNR. The SNR for this source is 32. The type A source has a radius of 12mm and lays on the region of descending SNR right to the peak. The SNR for this source is 20.

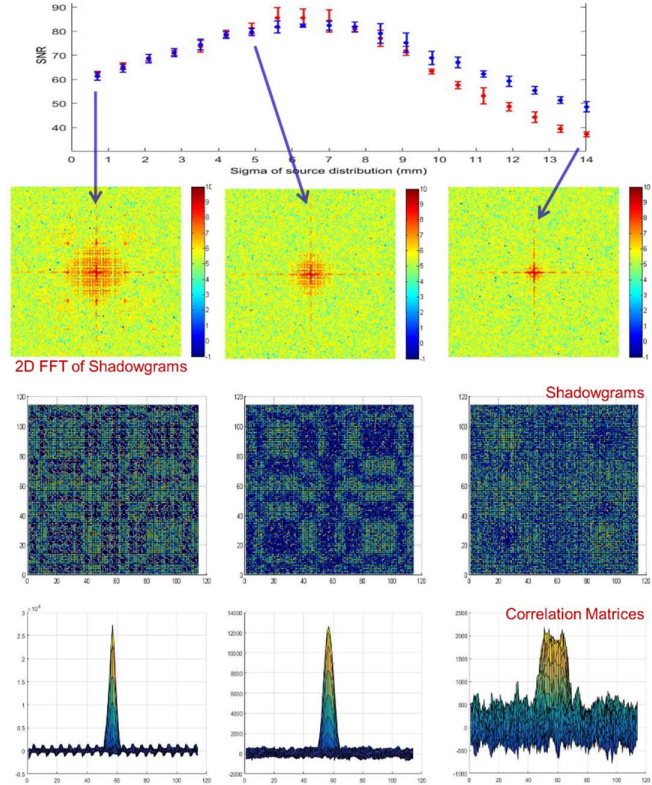

**Figure 3:** <u>SNR plot</u>: The Signal-to-Noise Ratio of the correlation matrix as function of the extent of the hot-spot, i.e. the standard deviation (sigma) of the source distribution. Blue: uniform distribution. Red: normal distribution. All the simulations performed with the mask 19R-1821 and the hot spot placed at 160mm distance from γ-camera. The error bars stand for the standard deviation of the results of five-time repeated simulation experiments.
<u>2D FFT plots</u>: In every column (top to bottom): the XY projection of the Fast Fourier Transform (FFT) of the shadowgram, the XY projection of the actual shadowgram, the YZ projection of the correlation matrix. In every row (left to right): Extended Tc-99m hot spot with 0.7mm, 5mm and 14mm standard deviation.

The standard deviation of the background of the correlation matrix, which includes the variation of the intrinsic noise, is slightly smaller for the source Type A than for the source Type B, when the mask 19R-1958 is used. For this combination of ideal source distance, spatial dimensions of hot-spots and element-pitch, the Type A source lays on the region of the peak of the SNR and the Type B source lays on the region of its left descending (Figure 3.SNR plots). The standard deviation of the background of the correlation matrices presented in Figure 4 is 333 for the Type A source and 337 for the Type B source, which is not a significant difference. However, the peaks and valleys of the intrinsic noise, like the ones appearing in Figure 2, are evident on the background of Figure 4.a, while they are eliminated on the background of Figure 4.b.

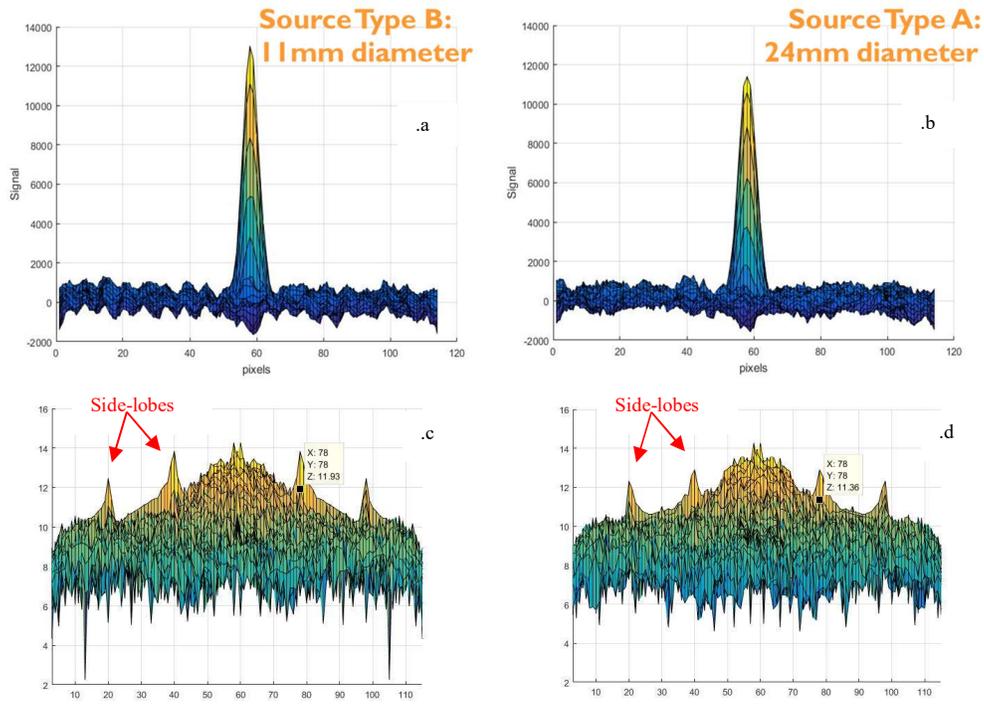

**Figure 4.** Top: Experimental correlation matrix for the $^{99m}$Tc Type B and Type A hot-spots, with the mask 19R-1958 and the hot spots placed at 308mm distance from the γ-camera. Bottom: Their Fast Fourier Transforms with the indication of a spatial frequency which corresponds to the "wave" structure and the side-lobes.

Following the existence of the two components in the background, the Fast Fourier Transform of the correlation matrices also has the two components: (i) the intrinsic and (ii) the random noise signal for each spatial frequency. The elimination of the peaks and valleys should appear as a reduction of the signal of their specific spatial frequencies, even though the random noise of these frequencies still exists. Indeed, the side-lobes from Figure 4.c to Figure 4.d are reduced and specifically a reduction of 5% (from 11.93 to 11.36) is evident for the signal with a spatial frequency correlated with the "wave" structure of the background noise. Therefore the intrinsic noise is reduced as the spatial dimensions of the source become larger from Figure 4.a to Figure 4.b.

### 3.2 The optimum geometry

The middle column of the 2-D FFT plots in Figure 3 indicates that there exists a geometrical setup that yields the maximum SNR of the correlation matrix. Specifically, for source dimension around 2 cm and with the mask 19R-1958, the source has to be placed around a distance of 30 cm from the γ-camera. If the mask 19R-1821 is used, the source must be placed around a distance of 16 cm from the γ-camera and its dimension has to be around 1 cm.

In nuclear medicine and intra-operative imaging the objects under investigation have certain spatial dimensions [8]. For example, the lymph nodes can be considered as spheres with diameter around 1 to 2 cm [9], [10]. Therefore, different geometrical configurations of the coded aperture γ-camera can be used for the imaging of the sentinel lymphs, or the thyroid residue, in order to achieve the best SNR possible.

### 3.3 The kernel filtering

The reduction of the intrinsic noise resulting from the increase of the dimensions of the hot-spot can be exploited to reduce this noise, also, in the correlation matrix of point-like sources, where this noise is the highest. Due to the periodicity of the peaks and valleys on the correlation matrix, scanning the whole surface of the correlation matrix with an appropriate kernel filter should reduce the intrinsic noise. Since the dimension of the peaks and valleys are proportional to the FWHM of the PSF of the correlation matrix, this filter (Figure 5) is chosen to be the cropped and normalized peak of the Auto-Correlation-Function (ACF) of each mask. For our mask rank, pixel size and geometry configuration, it is a matrix of 6 by 6 pixels. The result is the slight expansion of FWHM of the peak, as the peak undergoes a convolution with the normalized, cropped ACF and the significant reduction of the intrinsic noise on the correlation matrix, as one can see in Figure 5.

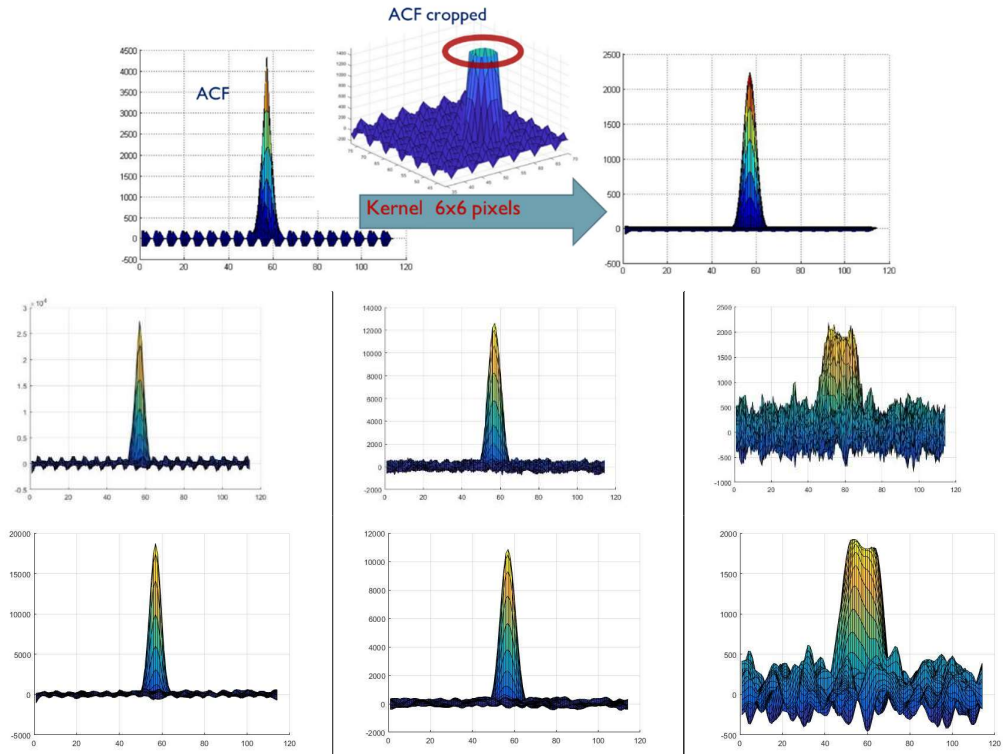

**Figure 5.** The kernel filter consists of the cropped normalized peak of the ACF. Its application on the correlation matrix reduces the intrinsic noise: Top: Kernel filter is applied on the ACF. In each column from left to right: Kernel filter is applied on the correlation matrix of extended Tc-99m hot spot with 0.7mm, 5mm and 14mm standard deviation.

In many cases of astronomy or security applications of coded apertures, the sources under investigation are considered point-like. The aforementioned filtering can be usefull for smoothing the background of the correlation matrix, which includes one or more peaks, related to point-like sources.

## 4. Conclusions

The proposed NTOT masks are simple in construction terms, less heavy than their conventional NTHT counterparts and have a high transparency ratio. This investigation of the maximization of the SNR for spatially extended radioactive hot-spots leads to the proposal of using configurations with certain geometrical characteristics, in order to reduce the intrinsic noise of the final image of the source. Moreover, for the case of point sources, a simple filtration method is proposed in order to achieve better SNR on the correlation matrices.

## Acknowledgments

This research is co-financed by Greece and the European Union (European Social Fund- ESF) through the Operational Programme «Human Resources Development, Education and Lifelong Learning 2014-2020» in the context of the project "Gamma Ray Tomographic Imaging for Medical and Industrial Applications", (MIS5007234).